\documentclass[prl,noshowkeys,superscriptaddress,twocolumn,
    preprintnumbers,amsmath,amssymb, showpacs]{revtex4}
\usepackage{graphicx,color}

\usepackage{amsmath,enumerate}

\newcommand{\tmop}[1]{\ensuremath{\operatorname{#1}}}

\newcommand{\ket}[1]{\left|#1\right\rangle}
\newcommand{\bra}[1]{\left\langle#1\right|}

\usepackage{amsmath}

\begin{document}

\title{How long can a quantum memory withstand depolarizing noise?}

\newcommand{\affilmpq}{\affiliation{Max-Planck-Institut f{\"{u}}r
Quantenoptik,
Hans-Kopfermann-Str.\ 1, D-85748 Garching, Germany.}}

\author{Fernando Pastawski}
\affilmpq
\author{Alastair Kay}
\affilmpq
\affiliation{Centre for Quantum Computation,
             Centre for Mathematical Sciences,
             University of Cambridge,
             Cambridge CB3 0WA, UK}
\author{Norbert Schuch}
\affilmpq
\author{Ignacio Cirac}
\affilmpq

\begin{abstract}
We investigate the possibilities and limitations of passive
Hamiltonian protection of a quantum memory against depolarizing
noise. Without protection, the lifetime of a single qubit is
independent of $N$, the number of qubits composing the memory. In
the presence of a protecting Hamiltonian, the lifetime increases at most logarithmically
with $N$. We construct an
explicit time-independent Hamiltonian which saturates this bound,
exploiting the noise itself to achieve the protection.
\end{abstract}


\maketitle

A cornerstone for the majority of applications in quantum information
processing is the ability to reliably store quantum information,
protecting it from the adversarial effects of the environment.
Quantum Error Correcting Codes (QECC) achieve this task by using a
redundant encoding and regular measurements which allow for the
detection, and subsequent correction, of errors
\cite{shor_fault-tolerant_1996,
aharonov_fault_1996,gottesman_1998,gottesman_2009}. An alternative approach uses
so-called \emph{protecting Hamiltonians}
\cite{kitaev_fault-tolerant_2003,bacon_operator_2006}, which
permanently act on the quantum memory and immunize it against small
perturbations. Its most attractive feature is that, in contrast to
QECC, it does not require any active action on the quantum memory,
just encoding and decoding operations at the time of storing
and retrieving the information. Whereas this approach may tolerate
certain types of perturbation \cite{dennis_topological_2002,
alicki_thermal_2008}, it is not clear if it is suitable in the
presence of depolarizing noise, something which QECC can deal
with.

In this Letter, we give a complete answer to this question. More
specifically, we consider the situation where a logical qubit 
is encoded in a set of $N$ qubits and allowed to evolve in the
presence of depolarizing noise and a protecting Hamiltonian. The
goal is to find the strategy delivering the longest lifetime,
$\tau$, after which we can apply a decoding operation and reliably
retrieve the original state of the qubit. By adapting ideas taken
from \cite{aharonov_limitations_1996}, it is established
that the lifetime cannot exceed $\log N$. An analysis of
the case in which no protecting Hamiltonian is used
presents markedly different behavior depending on whether we intend
to store classical or quantum information. Finally, we construct
a static protecting Hamiltonian that saturates the upper bound
$\tau\sim O(\log N)$. To this end, we first show how to achieve this
bound using a time--dependent Hamiltonian protection which
emulates QECC. We then introduce a clock gadget which exploits
the noise to measure time--similar to radiocarbon dating--thus
allowing us to simulate the previous time dependent protection
without explicit reference to time.

We consider a system of $N$ qubits, each of which is independently
subject to depolarizing noise at a rate $r$. 
The total state evolves as
\begin{equation}
\label{eq:depolarizingEvolution}
 \dot\rho(t) = -i[H(t),\rho(t)] -r\!\left[N\rho(t)-\!\sum_{n = 1}^N\tmop{tr}_n
 (\rho(t))\otimes\frac{\openone_n}{2}\right]\!.
\end{equation}
Furthermore, we shall allow for an arbitrary encoding of the
initial state as well as a final decoding procedure to
recover the information.

\emph{Protection limitations.---}Using purely Hamiltonian
protection, a survival time of $\tau\sim O(\log N)$ is the maximum
achievable. Intuitively, this is due to the fact that the
depolarizing noise adds entropy to the system at a constant rate,
while any reversible operation (i.e., Hamiltonian/unitary
evolution) will never be able to remove this entropy from the
system. Rather, in the best case, it can concentrate all the entropy in a subsystem, keeping the remaining part as clean as possible. This
entropic argument was first presented in
\cite{aharonov_limitations_1996}, where the authors investigated the
power of reversible computation (both classical and quantum)
subject to noise in the absence of fresh ancillas. To this end,
they considered the \emph{information content} ${I}(\rho)=N
-{S}(\rho)$ of the system, with $N$ the number of qubits and
$S(\rho) = - \tmop{tr}(\rho \log_2\rho)$. 
The information content upper bounds the number of classical bits
extractable from $\rho$, and thus ultimately also the number of
qubits stored in $\rho$. While the original statement about the
decrease of $I(\rho)$ is for discrete-time evolution,
it can be straightforwardly generalized to the continuous time
setting of Eq.~(\ref{eq:depolarizingEvolution}), where it states
that
\[
\frac{\mathrm{d}I(\rho)}{\mathrm{d}t} \le -r I(\rho)\ ,
\]
which yields that the information of the system is smaller than $\frac{1}{2}$
after a time $\frac{\ln 2N}{r}$.

Having established an upper bound for the scaling of $\tau$
with $N$, let us analyze whether this bound can be reached under
different circumstances. We start out with the simplest case where
we use no Hamiltonian protection (i.e., $H=0$) and show that
$\tau$ is independent of $N$; that is, no quantum memory effect
can be achieved. For that, we note that the effect of Eq.
(\ref{eq:depolarizingEvolution}) on each qubit may be expressed in
terms of a depolarizing channel
\[
\mathcal E_t(\rho)=\lambda(t)\rho+(1-\lambda(t))\frac{\openone}2
\]
where $\lambda(t)=e^{-rt}$. For $t\ge t_\mathrm{cl}$, where
$\lambda(t_\mathrm{cl})=\frac{1}{3}$, the resulting channel is
entanglement breaking \cite{horodecki_general_2003}. This remains
true if one incorporates encoding and decoding steps. 
It is simple to prove that for entanglement breaking qubit
channels, the average fidelity \cite{nielsen_2002} is upper-bounded by
$2/3$. 
Thus, the lifetime $\tau$ is smaller than $t_\mathrm{cl}=\ln{3}/r$, 
which is independent of $N$.

The previous argument does not apply to classical information, for
which an optimal storage time that is logarithmic in $N$ may be achieved.
The classical version of Eq. \ref{eq:depolarizingEvolution},
taking $H(t)\equiv 0$, is a system of $N$ classical bits subject
to bit flipping noise (each bit is flipped at a rate $r/2$). In
this case, encoding in a repetition code, and decoding via majority
voting, yields an asymptotically optimal information
survival time $O(\log N )$. 
Using optimal estimation \cite{massar_popescu_1995}
and this classical protocol in the encoding phase, the bound $2/3$
may be asymptotically reached for the quantum case.

\begin{figure}[t]
\begin{center}
    \includegraphics[width=0.80\columnwidth]{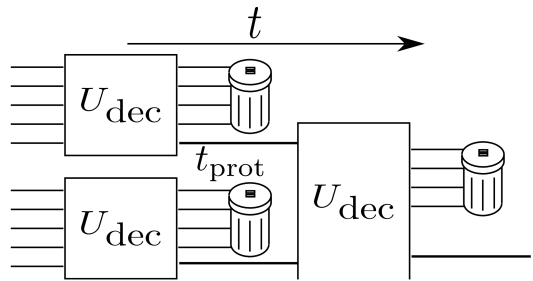}
\end{center}
\caption{ \label{fig:UnitaryDecoding} Decoding a nested QECC.  The
``discarded'' qubits carry most of
the entropy and are not used further.}
\end{figure}

\emph{Time dependent protection.---}We will now use the ideas of
QECC to build a simple circuit based model that reaches the upper
bound on the protection time. This model assumes that
unitary operations can be performed instantaneously, which is equivalent to having
a time--dependent protecting Hamiltonian with unbounded strength;
we will show how to remove both requirements later on. Instead of
using a repetition code, we encode the qubit to be protected in an
$l$ level concatenated QECC \cite{aharonov_fault_1996,gottesman_1998,gottesman_2009}
(i.e., $l$ levels of the QECC nested into each other), which
requires $N = d^l$ qubits, where $d$ is the number of qubits used
by the code. Each level of the QECC can provide protection for a
constant time $t_\mathrm{prot} < t_\mathrm{cl}$, and thus,
after $t_\mathrm{prot}$ one layer of decoding needs to be
executed. Each decoding consists of a unitary $U_{\tmop{dec}}$ on
each $d$-tuple of qubits in the current encoding level; after the
decoding, only one of each of the $d$ qubits is used further
(Fig.~\ref{fig:UnitaryDecoding}). The total time that such a
concatenated QECC can protect a qubit is given by
$t_\mathrm{prot}l=t_\mathrm{prot}\log_d N \sim O( \log N)$, as in the
classical case.

\emph{Time-independent protection.---}In the following, we show
that the same $\log N$ protection time which we can achieve using
a time-dependent protection circuit can also be obtained from a
time-independent protecting Hamiltonian. The basic idea of our
construction is to implement the time-dependent Hamiltonian
presented before in a time independent way. To this end, a clock
is built which serves as control. The time-independent version
performs the decoding gates conditioned on the time estimate
provided by the clock. In order to obtain a clock from
(\ref{eq:depolarizingEvolution}) with a time-independent $H$, we
will make use of the noise acting on the system: we add a number,
$K$, of ``clock qubits'' which we initialize to $\ket{1}^{\otimes
K}$ and let the depolarizing noise act on them. The behavior of
the clock qubits is thus purely classical; they act as $K$
classical bits initialized to $1$ which are being flipped at a rate $r/2$. Thus, the polarization $k$, defined by the
number of ``$1$'' bits minus the number of ``$0$'' bits has an 
average expected value of $\bar{k}(t)=Ke^{-rt}$ at  time $t$.
Conversely, this provides the time estimate
\begin{equation}
\label{eq:time-estimate}
\tilde{t}(k)=\tmop{min}\left(\frac{\ln(K/k)}{r},t_\mathrm{max}\right).
\end{equation}
Here, $t_\mathrm{max}$ is the maximum time for which we expect the
estimate to be reliable which depends on $K$ and the precision 
of the estimate, cf.~(\ref{eq:time-estimate-error}) below.
Particular realizations of this random process of bit flips can be
described by a polarization trajectory $k(t)$. \emph{Good trajectories} are
defined to be those such that
\begin{equation}
\label{eq:good-traj}
|k(t)-\bar k(t)| < K^{1/2+\epsilon}
\end{equation}
for all $0\le t\le t_\mathrm{max}$.
For appropriate parameters $t_\mathrm{max}$ and $0<\epsilon<\frac{1}{2}$, the following 
theorem states that almost all trajectories are good and can provide accurate 
time estimates.

\noindent \textbf{Theorem (Depolarizing clock):}
\emph{For $K\ge16$, \emph{good trajectories}
have a probability
\begin{equation}
\label{eq:goodtraj-prob}
P\left[k(t)\textrm{\emph{ good traj.}}\right]
 \geq  1-K\frac{r t_\mathrm{max}+\exp[-3K^{2\epsilon}\!/8]}{
    \exp[K^{2\epsilon}\!/8]}\ .
\end{equation}
Furthermore, for any good trajectory $k(t)$, the \emph{time
estimate} $\tilde t$ returned by the clock will differ from the real time $t$ by
at most
\begin{equation}
\label{eq:time-estimate-error}
 \frac{\delta}{2} := \frac{1}{rK^{1/2-\epsilon}} e^{rt_\mathrm{max}} \ge |\tilde t(k(t))-t|\ .
\end{equation}
} 
(For fixed $\delta$, this implies that $t_\mathrm{max}$ will scale
logarithmically with the number of qubits.)

Note that the theorem does not simply state that any time
evolution will be outside (\ref{eq:good-traj}) for an
exponentially small amount of time (which is
easier to prove), but that there is only an exponentially small
number of cases in which (\ref{eq:good-traj}) is violated \emph{at
all}.  Although the former statement would in principle suffice to
use the clock in our construction, the stronger version of the
theorem makes the application of the clock, and in particular the
error analysis, more transparent and will hopefully lead to
further applications of the clock gadget.

\textbf{Proof.}\ To prove the theorem, note that each of the bits
undergoes an independent exponential decay, so that the total
polarization is the sum of $K$ identical independent random variables.
We can thus use Hoeffding's inequality \cite{hoeffding_1963} to bound the
probability of finding a polarization far from the expected
average value $\bar{k} (t)$,
\begin{equation}
\label{eq:HoefdingIneq}
\Pr \left[ |k (t) - \bar{k} (t)| \geq K^{1/2+\epsilon} \right]
    \leq  2 e^{- \frac{K^{2\epsilon}}{2}}\ .
\end{equation}

This already implies that most of the trajectories violate
(\ref{eq:good-traj}) for no more than an exponentially small amount of time.
To see why (\ref{eq:HoefdingIneq}) implies that most trajectories
are good trajectories, we bound the average number of
times a trajectory leaves the region (\ref{eq:good-traj}) of
good trajectories. Since a non-good trajectory must leave (\ref{eq:good-traj}) at least once, it is also an upper bound on the probability of non-good
trajectories. Hence, it suffices to consider the average rate
$R(t)$ at which processes leave (\ref{eq:good-traj}), and
integrate over $t$ to obtain a bound on the probability of 
trajectories which are not good.

\begin{figure}[t]
\begin{center}
\includegraphics[width=0.9\columnwidth]{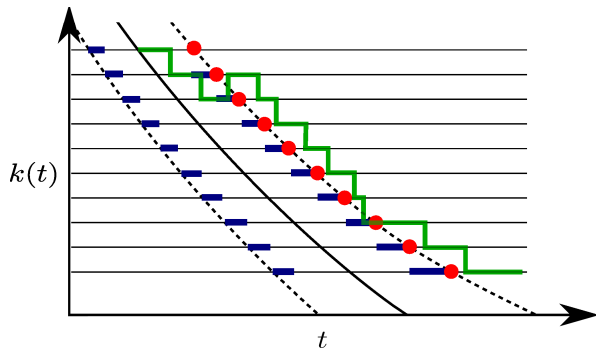}
\end{center}
\caption{ \label{fig:BadTrajectory} (color online) 
A step-like trajectory in green illustrates the two ways of leaving
region (\ref{eq:good-traj}) of good trajectories (dashed lines): either a spin flip can take the
polarization out of the marked region
(thick blue), or polarization may leave region
(\ref{eq:good-traj}) as time passes without a spin flip (red
dots).}
\end{figure}

The rate at which a process leaves the set of good trajectories
has two sources, as illustrated in Fig.~\ref{fig:BadTrajectory}:
First, the system can undergo a spin flip, thus leaving the region
defined by (\ref{eq:good-traj}) vertically (rate $R_v$), and
second, it can leave it horizontally if the time $t$ passes the
maximum time allowed by (\ref{eq:good-traj}) for the current value
$k(t)$ of the polarization (rate $R_h$). A vertical leave can
occur only if $|k(t)- \bar{k}(t)| \ge K^{1/2+\epsilon} -2 \ge
K^{1/2+\epsilon}/2$, provided $K\geq 16$ 
(a spin flip changes
$k(t)$ by $\pm2$). Eqn.~(\ref{eq:HoefdingIneq}) thus gives an
average bound
\[
R_v(t)\le
K r e^{-K^{2\epsilon}/8}\ .
\]
A horizontal leave can only occur at discrete times
extremizing (\ref{eq:good-traj}),
\[
t\in\mathcal T=\{t:\bar k(t)+K^{1/2+\epsilon} \in \mathbb N\}\ ,
\]
and the probability of a trajectory fulfilling
$k(t)=\bar k(t)+K^{1/2+\epsilon}$
may again be bounded using (\ref{eq:HoefdingIneq}), such that
\[
R_h(t)\le 2 e^{-K^{2\epsilon}/2}
    \sum_{\tau\in\mathcal T}\delta(t-\tau)\ .
\]
The inequality (\ref{eq:goodtraj-prob}) follows immediately by integrating
$R_h(t)+R_v(t)$ from $0$ to $t_\mathrm{max}$.

Assuming that $k(t)$ corresponds to a good trajectory, the
accuracy of the time estimate (\ref{eq:time-estimate}) may be
bounded by applying the mean value theorem to $\bar k$:
\[
|\tilde t(k(t))-t|=
    \frac{|\bar k(\tilde t(k(t)))-\bar k(t)|}{
    |\bar k'(t_\mathrm{interm})| }
    \le\frac{K^{\epsilon}}{r\sqrt{K}} e^{rt_\mathrm{max}}\ .
\]

\hspace*{\fill}$\square$

\textit{Clock dependent Hamiltonian.---}%
Let us now show how the decoding circuit can be implemented using the
clock gadget.  The circuit under consideration consists of the decoding
unitaries $U_\mathrm{dec}^{l,k}$ (decoding the $k$'th encoded qubit in
level $l$, acting on $d$ qubits each); after a time interval
$t_\mathrm{prot}$ (the time one level of the code can protect the qubit
sufficiently well), we perform all unitaries $U_\mathrm{dec}^{l,k}$ at the
current level $l$---note that they act on distinct qubits and thus commute.
Each of these unitaries can be realized by applying a $d$-qubit
Hamiltonian $H_\mathrm{dec}^{l,k}$ for a time
$t=t_\mathrm{dec}$. Thus, we have to switch on all the
$H_\mathrm{dec}^{l,\cdot}$ for $t\in [ t_l,t_l+t_\mathrm{dec}]$, where
$t_l=l\, t_\mathrm{prot}+(l-1)t_\mathrm{dec}$.

In order to control the Hamiltonian from the noisy clock, we define clock
times $k_{l,\mathrm{on}}=\lfloor\bar k(t_l)\rfloor$ and
$k_{l,\mathrm{off}}=\lceil\bar k(t_l+t_\mathrm{dec})\rceil$, and introduce a
time-independent Hamiltonian which turns on the decoding Hamiltonian for
level $l$ between $k\in[k_{l,\mathrm{on}},k_{l,\mathrm{off}}]$,
\begin{equation}
\label{eq:controlled-ham}
H=\sum_l \,
\left(H_\mathrm{dec}^{l,1}+\dots+H_\mathrm{dec}^{l,d^{L-l}}\right)
\otimes \Pi_l\ .
\end{equation}
The left part of the tensor product acts on the $N$ code qubits, the right part on the $K$ clock (qu)bits, and
\[
\Pi_l=\sum_{k=k_{l,\mathrm{on}}}^{k_{l,\mathrm{off}}}\sum_{w_x=(k+N)/2} \ket{x}\bra{x},
\]
where $x$ is an $N$-bit string with Hamming weight $w_x$. 
The initial state of the system is, as for the circuit
construction, the product of the encoded qubit in an $l$-level
concatenated code and the maximally polarized state
$\ket{1}^{\otimes K}$ on the clock gadget.

\emph{Error analysis.---}%
We now perform the error analysis for the protecting Hamiltonian
(\ref{eq:controlled-ham}). In order to protect the quantum
information, we will require that the error probability per qubit in use is bounded
by the same threshold $p^*$ after each decoding step is completed (i.e. at $t=t_l+t_\mathrm{dec}+\frac{\delta}{2}$). We will restrict to
the space of good trajectories, since we know from the clock
theorem that this accounts for all but an exponentially small
fraction, which can be incorporated into the final error probability.

We will choose $K$ large enough to ensure that the error
$\frac{\delta}{2}\ge|\tilde t-t|$ in the clock time satisfies
$\delta\ll t_\mathrm{prot},t_\mathrm{dec}$. In this way, we ensure
that the decoding operations are performed in the right order
\footnote{The noisy clock has the potential to run
backwards in time within its accuracy.} and with sufficient
precision. 
We may thus account for the following error sources between 
$t_l+t_\mathrm{dec}+\delta/2$ and $t_{l+1}+t_\mathrm{dec}+\delta/2$:

i) \emph{Inherited errors} from the previous rounds which could
not be corrected for. By assumption, these errors are bounded by
$p_\mathrm{inher}\le p^*$.

ii) Errors from the \emph{depolarizing noise} during the free
evolution of the system. The system is sure to evolve freely for a
time $t_\mathrm{prot}-\delta$, i.e., the noise per qubit is
bounded by
$p_\mathrm{evol}\le1-\exp[-r(t_\mathrm{prot}-\delta)]\le
r(t_\mathrm{prot}-\delta)$.

iii) Errors during the \emph{decoding}. These errors affect the
\emph{decoded} rather than the encoded system and stem from two
sources: On the one hand, the time the Hamiltonian is active has
an uncertainty $t_\mathrm{dec}\pm\delta$, which gives an error in
the implemented unitary of not more than
$\exp[\delta\|H_\mathrm{dec}^{k,l}\|]-1$. On the other hand,
depolarizing noise can act during the decoding for at most a time
$t_\mathrm{dec}+\delta$. In the worst case, noise on any of the
code qubits during decoding will destroy the decoded qubit, giving
an error bound $d(1-\exp[-r(t_\mathrm{dec}+\delta)]) \le
dr(t_\mathrm{dec}+\delta)$. Thus, the error on the decoded qubit
is
\[
p_\mathrm{dec}\le
\exp[\|H_\mathrm{dec}^{k,l}\|\delta]-1
+dr(t_\mathrm{dec}+\delta)\ .
\]
Since the noise is Markovian (i.e.~memoryless), the clock does not correlate its errors in time. In summary, the error after one round of decoding is at most
$B(p_\mathrm{inher}+p_\mathrm{evol})+p_\mathrm{dec}$, which we
require to be bounded by $p^*$ again. Here, $B(p)$ is a property
of the code, and returns the error probability of the decoded
qubit, given a probability $p$ of error on each of the original
qubits; for example, for the 5-qubit perfect QECC \cite{laflamme_miquel_paz_zurek_1996}, $B(p) \leq 10 p^2$. 

We will now show that it is possible to fulfill the required
conditions by appropriately defining the control parameters. 
First, we choose $p^* \leq 1/40$ to have the QECC 
\cite{laflamme_miquel_paz_zurek_1996} work well below threshold.
We may take $t_\mathrm{prot}:= \frac{p^*}{r}$ and $t_{\mathrm{dec}}:= \frac{p^*}{4dr}$.
To minimize imprecision in the implemented unitaries, the  
decoding Hamiltonians are chosen of minimal possible strength, 
$\|H_\mathrm{dec}^{k,l}\|\leq \frac{2\pi}{t_\mathrm{dec}}$.
Finally we take $\delta:=\frac{p^* t_\mathrm{dec}}{8\pi}$.
Inserting the proposed values in the derived bounds, it is straightforward to show that $B(p_\mathrm{inher}+p_\mathrm{evol}) + p_\mathrm{dec} < p^*$.

The number of code qubits required is $N:=d^l$, with $l:=\lceil
\frac{\tau}{t_\mathrm{prot}+t_\mathrm{dec}} \rceil$.
The required logarithmic clock lifetime $t_\mathrm{max}=\tau$ and the 
precision $\delta$
are obtained by taking $\epsilon=1/6$ 
and $K:=( \frac{2e^{r\tau}}{r\delta})^{3}$, by virtue of
Eq.~(\ref{eq:time-estimate-error}) of the
clock theorem.
For any fixed $r$ and $p^*$, this allows a lifetime 
$\tau \sim O(\log (N+K))$.

\emph{Conclusions.---}In this paper, we have considered the
ability of a Hamiltonian to protect quantum information from
decoherence. While without a Hamiltonian, quantum information is
destroyed in constant time, the presence of time-dependent
control engenders protection for logarithmic time, which is
optimal. As we have shown, the same level of protection can
be attained with a time-independent Hamiltonian. The construction introduced a noise-driven clock which allows a
time dependent Hamiltonian to be implemented without explicit reference to time.

Since depolarizing noise is a limiting case of local noise models, it is expected that the time-independent Hamiltonian developed here can be tuned to give the same degree of protection against weaker local noise models, although these models may admit superior strategies. For instance, noise of certain forms (such as dephasing) allows for storage of ancillas, potentially yielding a linear survival time by error correcting without decoding. In the case of amplitude damping noise, the noise itself distills ancillas so that the circuit can implement a full fault-tolerant scheme, which gives an exponential survival time, assuming that one can redesign the clock gadget to also benefit from these properties.

Whether the same degree of
protection can be obtained from a Hamiltonian which is local on a 2D or 3D
lattice geometry remains an open question~\footnote{A first step is to incorporate the notion of boundedness. By controlling each decoding unitary in a given round from a different clock (which does not affect the scaling properties), a constant bound to the sum of Hamiltonian terms
acting on any given finite subsystem can be shown. }. However, intuition
suggests this might be impossible; the crucial point in reversibly protecting
quantum information from depolarizing noise is to concentrate
the entropy in one part of the system. Since the speed of
information (and thus entropy) transport is constant due to the
Lieb-Robinson bound \cite{lieb_robinson_1972}, the rate at which 
entropy can be removed from a given volume is proportional to its
surface area, while the entropy increase goes as the
volume. It thus seems impossible to remove the entropy
sufficiently quickly, although this argument is not fully
rigorous, and the question warrants further
investigation.

We acknowledge the DFG for support (projects MAP and
FOR635), and the EU (QUEVADIS and SCALA). ASK is also supported by Clare
College, Cambridge.

\end{document}